\documentclass[aps,prl, 
twocolumn,
groupedaddress]{revtex4-2}

\usepackage{dsfont}
\usepackage{graphicx}
\usepackage{dcolumn}
\usepackage{braket}
\usepackage{hyperref}
\usepackage{amsmath}
\usepackage{amssymb}
\usepackage{tikz}

\begin{document}


\title{\textbf{Deconfined quantum criticality in a frustrated Haldane chain with single-ion anisotropy} 
}%

\author{Niels T. Pronk}
\author{Bowy M. La Rivi\`ere}%
 \email{Contact author: b.m.lariviere@tudelft.nl}
\author{Natalia Chepiga}%
\affiliation{ Kavli Institute of Nanoscience, Delft University of Technology, Lorentzweg 1, 2628CJ Delft, The Netherlands }


\date{\today}

\begin{abstract}
We report a phase diagram of the antiferromagnetic spin-1 chain with nearest-neighbor Heisenberg and three-site interactions in the presence of single-ion anisotropy. We show that the Gaussian and Ising transitions that separate the topological Haldane phase from the two anisotropic phases eventually fuse into a higher symmetry point characterized by the Wess-Zumino-Witten (WZW) SU(2)$_2$ critical theory providing a lattice realization of the conformal embedding. On the other side of the WZW multi-critical point, the Ising critical line reappears together with the eight-vertex transition. This transition is a one-dimensional realization of a deconfined quantum criticality separating the dimerized and Ising antiferromagnetic phases - two ordered phases with incompatible order parameters.
\end{abstract}


\maketitle

{\bf Introduction.} 
Quantum phase transitions beyond the Landau-
Ginzburg-Wilson-Fisher (LGWF) paradigm attracted a lot of attention in the last two decades \cite{sachdevQuantumPhaseTransitions2011,senthil_book, Senthil_2005,PhysRevX.7.031051,PhysRevX.9.021034}. 
The paradigmatic example is deconfined quantum criticality between the valence-bond singlet (VBS) and magnetic  N\'eel phase  with two incompatible order parameters \cite{senthil_DQC,PhysRevB.70.144407, PhysRevLett.98.227202,PhysRevX.5.041048}. 
The appearance of a continuous transition between these phases violates the LGWF-paradigm and is accompanied by the emergence of fractionalized degrees of freedom.
While the problem of deconfined criticality originates and is often attributed to 2+1D \cite{senthil_book,Sandvik_2020,PhysRevLett.133.166702}, here we address the problem in the lower 1+1D dimensional context and show that it can be traced back to the rare yet well-understood conformal transition in the eight-vertex universality class \cite{BAXTER1972323,BAXTER1972193,8vertex}.

\begin{figure}[t]
    \centering
    \includegraphics[width=1.0\linewidth]{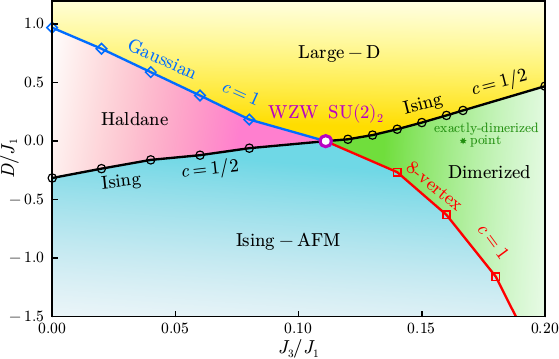}
    \caption{
    Phase diagram of the spin-1 $J_1-J_3$ chain with a single-ion anisotropy $D(S_i^z)^2$ defined in Eq. \eqref{eq:Hamiltonian}. 
    It contains four gapped phases: topologically non-trivial Haldane phase, Ising-antiferromagnetic (AFM) phase, the dimerized phase, and the trivial large-D phase. 
    First and second pair of phases are separated by two Ising transitions characterized by a central charge $c=1/2$.
    The transition between the Haldane and large-D phase is the Gaussian topological transition with $c=1$. 
    The transition between the dimerized and Ising-AFM phase realizes a deconfined quantum criticality and belongs to the eight-vertex universality class - another realization of the Gaussian $c=1$ transition taking place between two ordered phases. 
    The multi-critical point is described by the Wess-Zumino-Witten (WZW) SU(2)$_2$ critical theory.
    }
    \label{fig:phase diagram}
\end{figure}

In this Letter, we report the appearance of the 1D deconfined quantum phase transition in a frustrated Haldane chain induced by a single-ion anisotropy. 
We study the following microscopic spin-1 Hamiltonian:
\begin{multline} \label{eq:Hamiltonian}
H=\sum_i J_1 \mathbf{S}_{i}\cdot \mathbf{S}_{i+1} +J_3 [\left(\mathbf{S}_{i-1}\cdot\mathbf{S}_{i}\right)\left(\mathbf{S}_{i}\cdot\mathbf{S}_{i+1}\right) + \mathrm{H.c}.]\\ 
        + D\left(S_i^z\right)^2,
\end{multline}
where $J_1$ is an antiferromagnetic Heisenberg interaction, that without loss of generality we set to $J_1=1$, and $D$ is a single-ion anisotropy breaking the SU(2) symmetry down to U(1).
The three-site $J_3$ term appears along with the biquadratic and next-nearest-neighbor interactions in next-to-leading order in the strong coupling expansion of the two-band Hubbard model \cite{michaudAntiferromagneticSpinSChains2012}. 
The isotropic $J_1-J_3$ model is a generalization of the Majumdar-Ghosh point \cite{majumdarNextNearestNeighborInteractionLinear1969} to higher spin $S$ realizing the exactly-dimerized state at $J_3=J_1/\left[4S(S+1)-2\right]$ \cite{michaudAntiferromagneticSpinSChains2012}. 
This point belongs to an extended dimerized phase with spin-1 singlets occupying every other bond - the 1D realization of the VBS phase. 
On the other hand, for small values of $J_3$ the system is in the Haldane phase \cite{HALDANE1983464} - topologically non-trivial gapped phase with emergent spin-1/2 edge states \cite{TKennedy_1990,PhysRevLett.65.3181}. 
The transition between the Haldane and dimerized phases takes place at $J_3\approx0.111$ and belongs to the Wess-Zumino-Witten (WZW) SU(2)$_2$ universality class \cite{affleckCriticalTheoryQuantum1987b,michaudAntiferromagneticSpinSChains2012}, indicated in Fig.\ref{fig:phase diagram}. 
A phase transition of the same nature appears as the exactly solvable Takhtajan-Babujian  critical point, separating the Haldane and dimerized phases in the bilinear-biquadratic spin-1 chain \cite{babujian_1982,takhtajan_1982}.

The single-ion anisotropy eventually destroys the Haldane phase and leads to either the trivial large-D phase for $D>0$ that, up to quantum fluctuations, corresponds to a uniform state of the type $|... 0,0,0,0, ... \rangle$; 
or for $D<0$ to a state that essentially excludes the $S^z_i=0$ degrees of freedom, reducing the system to an Ising anti-ferromagnet (AFM) with two ground states $|... 1,-1,1,-1 ... \rangle$ and $|... -1,1,-1,1 ... \rangle$ \cite{PhysRevB.28.3914,PhysRevB.67.104401,PhysRevB.34.6372,PhysRevB.104.024409,PhysRevB.84.054451}. 
For $J_3=0$ the Haldane phase is separated from the large-D phase by a topological Gaussian transition characterized by a central charge $c=1$ \cite{PhysRevB.67.104401,ercolessi,PhysRevB.84.220402,haldaneSpontaneousDimerization$Sfrac12$1982,nomuraCriticalProperties11994}. 
The phases on either side of this transition have no local order, and we will refer to this transition as the disordered Gaussian. 
For $D<0$ the transition between the Haldane phase and the Ising-AFM phase is in the Ising universality class \cite{PhysRevB.34.6372,PhysRevB.67.104401}. 

We study the ground-state phase diagram of the model of Eq.\ref{eq:Hamiltonian} numerically with the state-of-the-art Density Matrix Renormalization Group (DMRG) algorithm (see Supplemental Material \cite{SM} for details) \cite{whiteDensityMatrixFormulation1992a, whiteDensitymatrixAlgorithmsQuantum1993a}.

{\bf WZW SU(2)$_2$ multicritical point.} 
Upon increasing the three-site coupling $J_3$ we established that the Ising and disordered Gaussian transitions persist (see  Supplemental Material \cite{SM}) and approach each other, until they eventually fuse into the WZW SU(2)$_2$ multi-critical point at $D=0$.
This fusion is a lattice realization of the conformal embedding of a $\mathbb{Z}_2$ Ising parafermion with a central charge $c=1/2$ and a $c=1$ critical boson into the WZW SU(2)$_2$ critical theory with $c=3/2$ \cite{zamolodchikovNonlocalParafermionCurrents1985b}. 
One branch of this embedding has been observed in spin-1 chains - an isolated Ising transitions emerging in the presence of next-nearest-neighbor (NNN) interactions \cite{chepigaDimerizationTransitionsSpin12016a,chepigaCommentFrustrationMulticriticality2016a}. 
The latter eventually destroys the WZW transition and stabilizes the topologically-trivial NNN-Haldane phase separated from the dimerized one by a non-magnetic Ising transition. 
However, the Haldane and NNN-Haldane phases are separated by a first order transition and the $c=1$ branch was not observed \cite{chepigaDimerizationTransitionsSpin12016a,chepigaCommentFrustrationMulticriticality2016a}.
In the present case, by explicitly destroying the SU(2) symmetry, we can track both the Gaussian and Ising transitions to their fusion point where the SU(2) symmetry is restored. 
Quite remarkably, beyond the multicritical point, we detect a re-appearance of the Ising transition along with another realization of the Gaussian transition.
Below we present both of them in detail. 

\begin{figure}
    \centering
    \includegraphics[width=0.49\textwidth]{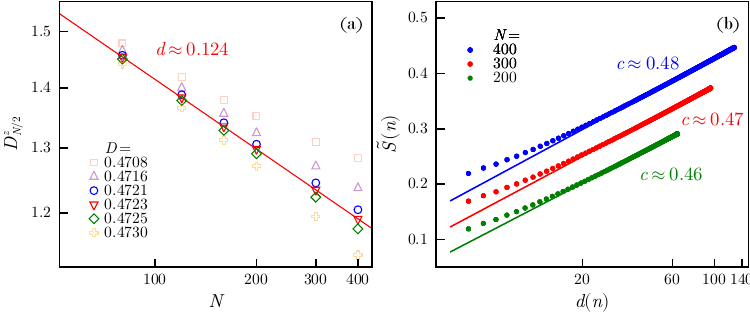}
    \caption{Numerical evidence for the Ising transition between the dimerized and trivial large-$D$ phases for $J_3 =0.2$.
    (a) Finite-size scaling of the mid-chain dimerization $D_{N/2}^z$ in a log-log plot.
    The transition is associated with the separatrix (red triangles) with $D_c\approx0.4723$, its slope $d\approx 0.124$ (red line) is in excellent agreement with the theory prediction for the scaling dimension $d=1/8$ for Ising transition. 
    (b) Scaling of the reduced entanglement entropy $\tilde{S}_N(n)$ with the conformal distance $d(n)$ at the critical point $D_c$ in a semi-log scale. 
    The central charge extracted with the Calabrese-Cardy formula in Eq.\eqref{eq:entanglement_entropy} agrees within 5\% with the CFT predictions $c=1/2$. 
    The results for $N=300$ and $200$ sites are shifted for visual clarity. 
    }
    \label{fig: Ising D>0}
\end{figure}

{\bf The Ising transition for $D>0$} separates the trivial large-D phase from the dimerized one.
We numerically locate this transition by performing a finite-size scaling of the local order parameter - the dimerization:
\begin{equation}\label{eq:dimerization_order_parameter}
    D_i^z = \left|\left\langle \hat{S}_i^z \hat{S}_{i+1}^z - \hat{S}_{i+1}^z \hat{S}_{i+2}^z \right\rangle \right|.
\end{equation}
According to boundary conformal field theory (bCFT) at the critical point, the middle-chain dimerization scales with the system size as $D^z_{N/2} \propto N^{-d}$ \cite{chepigaDimerizationTransitionsSpin12016a}. 
We associate the critical point with a straight line in a log-log scale of the scaling of the dimerization as a function of $N$; the slope of this separatrix corresponds to the scaling dimension $d$. 
In Fig.\ref{fig: Ising D>0}(a) we show an example of such a scaling for $J_3=0.2$.
The numerically extracted scaling dimension is in excellent agreement with the CFT prediction $d=1/8$ for the Ising transition. 

At the critical point, we extract the central charge $c$. 
For this we compute the entanglement entropy $S(n)$ from the reduced density matrix $\rho_n$ as $ S(n) = - \mathrm{Tr} \rho_n \ln \rho_n $. 
Then we calculate the reduced entanglement entropy by removing Friedel oscillations \cite{PhysRevLett.96.100603,capponiQuantumPhaseTransitions2013}:
$\tilde{S}(n) = S(n) - \zeta \langle \hat{S}^z_{n} \hat{S}^z_{n+1} \rangle$, where $\zeta$ is a non-universal constant.
Finally, we fit $\tilde{S}(n)$ to the Calabrese-Cardy formula \cite{calabreseEntanglementEntropyQuantum2004}:
\begin{equation}
    \label{eq:entanglement_entropy}
    \tilde{S}(n) = \frac{c}{6} \ln d(n) + s_1 + \ln(g),
\end{equation}
where $d(n) =\frac{2N}{\pi}\sin\left( \frac{\pi n}{N}\right)$ is the conformal distance, and $\ln g$ and $s_1$ are non-universal constants.
We show a typical example of $\tilde{S}(n)$ as a function of $\ln d(n)$ for three chain lengths in Fig.\ref{fig: Ising D>0}(b). 
The numerically extracted cental charge agrees, within a few percent, with the CFT prediction $c=1/2$ for the Ising universality class.

{\bf Eight-vertex transition.} 
Now let us discuss the transition between the dimerized and Ising-AFM phases. 
These phases are characterized by spontaneously broken $\mathbb{Z}_2$ symmetries that are incompatible with each other: the dimerized phase by the local bond-order operator $D^z_i$ defined in Eq.\ref{eq:dimerization_order_parameter}, while Ising anti-ferromagnet has site ordering captured with the local order parameter $m^z_i$:
\begin{equation}\label{eq: AFM order parameter}
    m_i^z = \left| \langle \hat{S}_i^z - \hat{S}_{i+1}^z\rangle \right| .
\end{equation}

\begin{figure}
    \centering
    \includegraphics[width=1.0\linewidth]{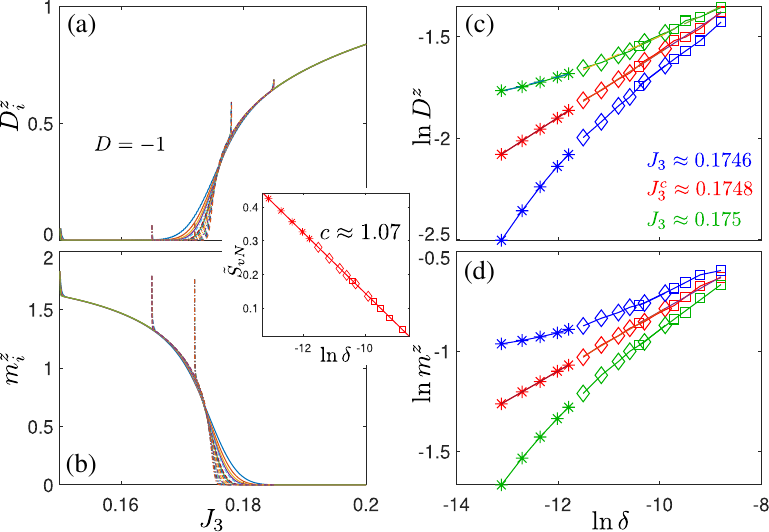}
    \caption{
    (a)-(b) Profiles of the local  (a)  dimerization $D^z_i$ and (b) magnetic alternation $m^z_i$ as a function of local value of $J_3(i)$ linearly changing along a chain. (c), (d) scaling of these order parameters with the gradient $\delta=J_3(i+1)-J_3(i)$ in a log-log scale. 
    The transition appears as a separatrix in {\it both} scaling plots indicating a direct transition between the two ordered phases. 
    We present results for $D=-1$ and for three intervals of $J_3$:  $[0.14,0.21]$ (solid lines, squares), $[0.165,0.185]$ (dashed, diamonds),  $[0.172,0.178]$ (dash-dotted, stars) and for various system sizes up to $N=3000$. In order to break the degeneracy and to measure magnetization $m^z_i$ locally we use polarized boundary condition at the edge inside the AFM-Ising phase.
    Inset: Scaling of the entanglement entropy with the gradient rate $\delta$ at $J_3^c\approx0.1748$. The extracted value of the central charge is in a reasonable agreement with $c=1$ critical theory.
    }
    \label{fig:scandmrg}
\end{figure}

We start by locating the transition using the scan-DMRG algorithm \cite{scan_DMRG}, which appears to be a perfect tool for capturing the transition between two ordered phase at once (see Supplemental Material \cite{SM} for details).
In scan-DMRG the external parameter, here we use $J_3$, gradually changes along the chain such that two edges are located inside the gapped phases on two different sides of the transition of interest.
We break translation symmetry of the Ising-AFM region by explicitly polarizing the left boundary along the $S^z$ direction while that of the dimerized part is naturally broken with an open right boundary.
In Fig.\ref{fig:scandmrg}(a)-(b) we present the profile of the two local order parameters for $D=-1$ with various start and end values of $J_3$, and various system sizes ranging from 400 to 3000 sites.
The transition out of an ordered phase appears as a separatrix in the log-log scaling of the order parameters as a function of the gradient step $\delta=J_3(i+1)-J_3(i)$ \cite{scan_DMRG}.
From Fig.\ref{fig:scandmrg}(c)-(d) one can easily see that both order parameters lead to the same critical point $J_3\approx 0.1748\pm 10^{-4}$, suggesting with a high confidence a direct transition between the two ordered phases.

Knowing the accurate location of the critical point, we extract the central charge from the scaling of the reduced entanglement entropy $\tilde{S}(n)$ with the gradient $\delta$ \cite{scan_DMRG}. 
To remove Friedel oscillations, we follow the same procedure as described above for the conventional DMRG algorithm.
The extracted value of the central charge $c\approx1.07$ (inset Fig.\ref{fig:scandmrg}) points to one of the $c=1$ critical theories.

\begin{figure}
    \centering
    \includegraphics[width=0.45\textwidth]{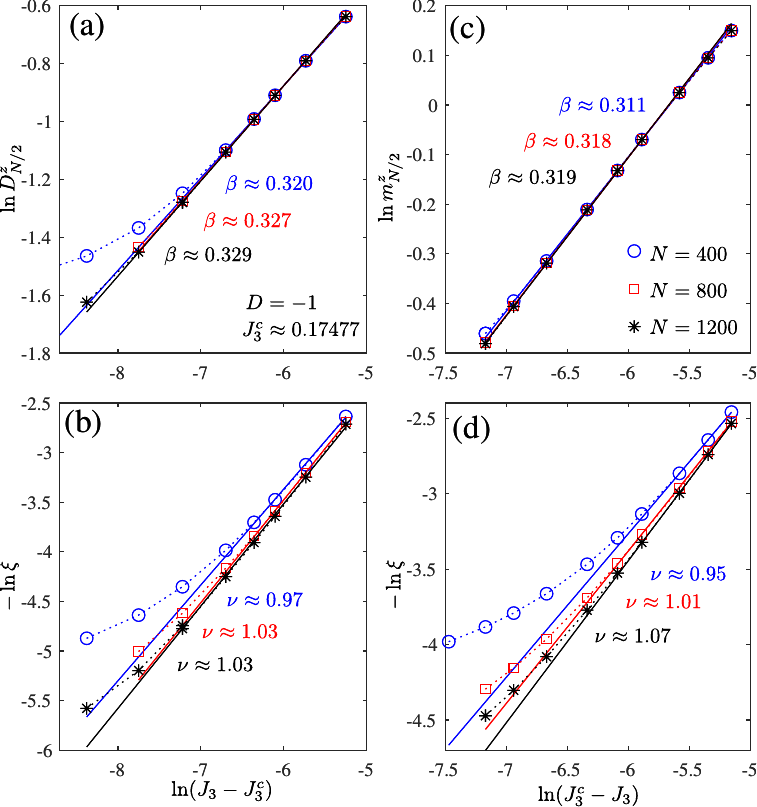}
    \caption{Scaling of the two local order parameters, (a) $D^z_{N/2}$ and (c) $m^z_{N/2}$, and (b), (d) the correlation length $\xi$ as a function of distance to the critical point $J_3^c$ identified with the scan-DMRG as shown in Fig.\ref{fig:scandmrg}.
    The results are taken inside (a)-(b) the dimerized phase, and (c)-(d) the Ising-AFM phase.
    Symbols are our numerical data, solid lines are linear fit in the log-log scale, dotted lines are guide to eyes.
    }
    \label{fig:nu_beta}
\end{figure}

We extract critical exponents $\beta$, associated with each order parameter, and the critical exponent $\nu$, controlling the divergence of the spin-spin correlation length, using the conventional finite-size DMRG algorithm.
We do this independently on two sides of the transition, assuming, however, a single critical point with its location determined with scan-DMRG.
Examples of the scaling for both order parameters and the correlation length are presented in Fig.\ref{fig:nu_beta}. 
While in the dimerized phase the finite-size effects become negligible for $N=1200$, the results extracted in the Ising-AFM still show stronger finite-size dependence (for instance, the point at which the curve starts to deviate from a linear scaling for a given system size). 
This effect has been reported before for other models featuring eight-vertex criticality, including the integrable one \cite{8vertex}.
We estimate the errors from the fits to be within $5-7\%$ due to the parameter window selected for each fit.

\begin{figure}
    \centering
    \includegraphics[width=0.49\textwidth]{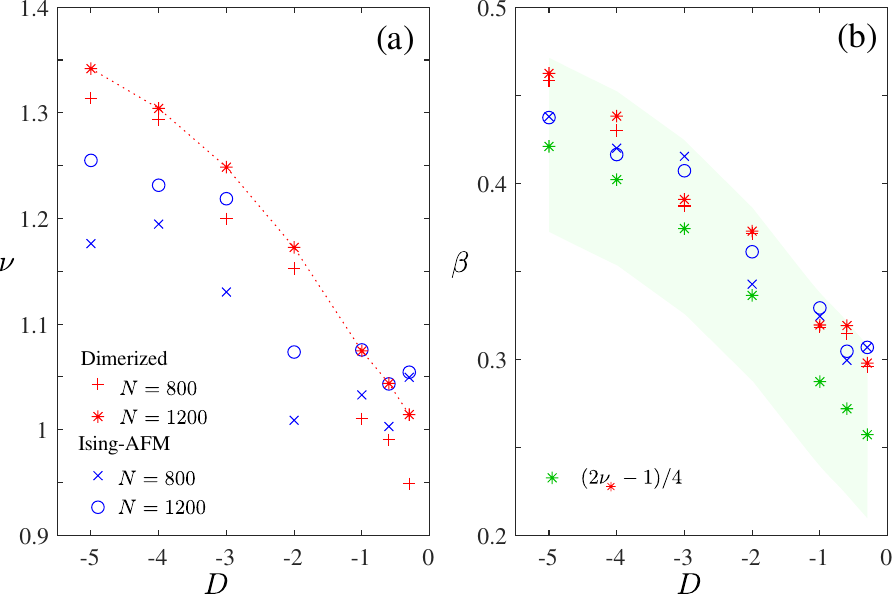}
    \caption{
    Critical exponents (a) $\nu$ and (b) $\beta$ (b) along the deconfined critical line.
    Red (blue) symbols correspond to the exponents extracted in the dimerized (Ising-AFM) phase, that are in a reasonable agreement with each other supporting the direct transition between the two phases.
    Both exponents, $\beta$ and $\nu$, show a strong dependence on the single-ion anisotropic strength $D$.
    Red dotted line is a guide to eyes.
    (b) Green stars show $\beta_\mathrm{eight-vertex}=(2\nu-1)/4$ with $\nu$ obtained in the dimerized phase with $N=1200$ (red stars in (a)).
    Based on the scattering of data points in (a) we roughly estimate numerical uncertainty in $\nu$ to be about $10\%$, indicated as a green shaded area in (b). $\beta_\mathrm{eight-vertex}$ (green stars) and $\beta$ extracted from the order parameters (red and blue symbols) are in good agreement and provide solid evidence that this transition belongs to the eight-vertex universality class.
    }
    \label{fig:8vertex}
\end{figure}

In Fig.\ref{fig:8vertex} we summarize the numerically extracted critical exponents along the transition line.
The ones obtained in the dimerized (red) and Ising-AFM (blue) phases are in reasonable agreement with each other, further supporting our earlier conclusion regarding the direct nature of the transition. 
But more importantly, we see that $\nu$ and $\beta$ systematically change along the critical line, behaviour typical for deconfined quantum criticality. 
On the other hand, varying critical exponents signal a {\it weak} universality class - individual critical exponents are not fixed to a universal value, but only certain combinations of them.
One of the widely known examples is the Ashkin-Teller critical theory with continuously varying critical exponents ranging from $\nu=1$ in the limit of two decoupled Ising chains to $\nu=2/3$ at the four-state Potts point, but characterized by the universal scaling ratio $\beta/\nu=1/8$ \cite{difrancescoConformalFieldTheory1997,PhysRevB.24.5229,PhysRevB.91.165129}. 
In contrast, for the integrable XYZ-model with $J_x=-J_z$ the eight-vertex universality class controlled by a single parameter $\rho = \mathrm{acos}[(1- J_y/J_x)/(1 + J_y/J_x)]$: $\nu = \pi/(2\rho)$ and $\beta = (\pi- \rho)/(4\rho)$ \cite{Den_Nijs,BAXTER1972323,BAXTER1972193}.
For the non-integrable case $\rho$ is generically not known, but the relation it imposes on $\nu$ and $\beta$ is universal \cite{8vertex,10.21468/SciPostPhys.14.6.152,PhysRevB.108.054509}. 
In other words, by substituting $\rho=\pi/(2\nu)$ in the formula for $\beta$, we obtain the expression 
\begin{equation}
    \beta=\frac{2\nu-1}{4},
\end{equation}
that is expected to hold everywhere along the eight-vertex critical line.
In Fig.\ref{fig:8vertex}(b) we compare $(2\nu-1)/4$ shown in green with the actual critical exponents $\beta$ extracted numerically with the two local order parameters.
Given that for eight-vertex criticality the critical exponent $\beta$ can range from 0 to infinity, the agreement that we have found in Fig.\ref{fig:8vertex} is very convincing.

{\bf Discussion}. 
Let us now support the appearance of the eight-vertex critical line by symmetry arguments. 
Baxter's original model \cite{BAXTER1972323,BAXTER1972193} can be rigorously mapped to the interacting Kitaev chain, where the eight-vertex criticality appears as the transition between a phase with spontaneously broken parity and an Ising-AFM phase with broken translation symmetry \cite{8vertex}.
Similarly to the present case, both gapped phases spontaneously break $\mathbb{Z}_2$ symmetries incompatible with each other.
Here, however, instead of parity the dimerized phase leads to a $\mathbb{Z}_2$-broken bond order still incompatible with an Ising-AFM, though measurable locally with the dimerization parameter. 
A similar transition is realized in the spin-1/2 XXZ chain \cite{haldaneSpontaneousDimerization$Sfrac12$1982,nomuraCriticalProperties11994} and ladders \cite{oginoContinuousPhaseTransition2021,oginoSymmetryprotectedTopologicalPhases2021,fontaineSymmetryTopologyDuality2024} separating the N\'eel and dimerized phases with the critical theory being Gaussian and the same relation between critical exponents $\beta$ and $\nu$.
In this respect, we expect the effective critical theory of the transition line here to be Gaussian as well, and thus eight-vertex criticality appears as a special realization of the Gaussian critical theory with relevant operators that lead to two spontaneously broken incompatible $\mathbb{Z}_2$ orders.



The appearance of deconfined quantum criticality is always associated with non-trivial topological properties of the system. 
Let us argue that Ising-AFM phase, despite being an extremely simple ordered phase, can be regarded as topologically non-trivial.
Firstly, the finite-size splitting of the two ground states vanishing exponentially fast with the system size (see Supplemental Material \cite{SM}) is compatible with the emergence of two Majorana edge modes (in analogy with the parity-broken phase of the transverse field Ising and Kitaev models). 
Secondly, the Ising transition separating the Haldane phase from the Ising-AFM is non-topological, implying that both phases have to be topologically non-trivial.
By contrast, the eight-vertex transition is topological: it connects a topologically-trivial dimerized phase, to, therefore, a topologically non-trivial Ising-AFM. 

We expect a similar deconfined quantum criticality of the eight-vertex type to appear in the bilinear-biquadratic model with single-ion anisotropy.
However, in the absence of the exactly dimerized point, the extent of the dimerized phase is relatively small \cite{weyrauch}, making the potential study of the transitions out of it quite challenging.

Remarkably, the previously reported results on deconfined quantum criticality in the spin-1/2 chain \cite{Roberts_2019} would be consistent with the eight-vertex universality class identified here.

The realization of the eight-vertex transition due to single-ion anisotropy has put the problem of deconfined criticality in 1+1D into a new perspective.
First, there are many Ni materials with a significant single-ion anisotropy, including NENP \cite{PhysRevB.50.9174}, NINAZ \cite{PhysRevB.53.15004}, and NDMAP \cite{PhysRevB.63.104410}.
The strong dimerization resulting from the three-body interaction makes the interaction considered here the best starting point for the realization of 1D deconfined quantum criticality.
However, we would like to emphasize that any competing interaction(s) that would stabilize a dimerized state, including three-body and biquadratic interactions, would be a valid candidate.
Second, the versatility of the order arising due to frustration (trimerized, plaquette, etc.) naturally opens a possibility for generalizing spontaneously broken symmetry to $\mathbb{Z}_k$ while preserving the topological component in the problem through the Haldane phase.

\begin{acknowledgments}
We would like to thank Hong Hao Tu for useful comments on the Gaussian transition. NC thanks to Frederic Mila for insightful discussions and inspiring work on a related subject. This research has been supported by Delft Technology Fellowship. Numerical simulations have been performed with the Dutch national e-infrastructure with the support of the SURF Cooperative and at the DelftBlue HPC.
\end{acknowledgments}

\bibliography{Letter, comment}

\providecommand{\noopsort}[1]{}\providecommand{\singleletter}[1]{#1}%
\begin{thebibliography}{56}%
\makeatletter
\providecommand \@ifxundefined [1]{%
 \@ifx{#1\undefined}
}%
\providecommand \@ifnum [1]{%
 \ifnum #1\expandafter \@firstoftwo
 \else \expandafter \@secondoftwo
 \fi
}%
\providecommand \@ifx [1]{%
 \ifx #1\expandafter \@firstoftwo
 \else \expandafter \@secondoftwo
 \fi
}%
\providecommand \natexlab [1]{#1}%
\providecommand \enquote  [1]{``#1''}%
\providecommand \bibnamefont  [1]{#1}%
\providecommand \bibfnamefont [1]{#1}%
\providecommand \citenamefont [1]{#1}%
\providecommand \href@noop [0]{\@secondoftwo}%
\providecommand \href [0]{\begingroup \@sanitize@url \@href}%
\providecommand \@href[1]{\@@startlink{#1}\@@href}%
\providecommand \@@href[1]{\endgroup#1\@@endlink}%
\providecommand \@sanitize@url [0]{\catcode `\\12\catcode `\$12\catcode `\&12\catcode `\#12\catcode `\^12\catcode `\_12\catcode `\%12\relax}%
\providecommand \@@startlink[1]{}%
\providecommand \@@endlink[0]{}%
\providecommand \url  [0]{\begingroup\@sanitize@url \@url }%
\providecommand \@url [1]{\endgroup\@href {#1}{\urlprefix }}%
\providecommand \urlprefix  [0]{URL }%
\providecommand \Eprint [0]{\href }%
\providecommand \doibase [0]{https://doi.org/}%
\providecommand \selectlanguage [0]{\@gobble}%
\providecommand \bibinfo  [0]{\@secondoftwo}%
\providecommand \bibfield  [0]{\@secondoftwo}%
\providecommand \translation [1]{[#1]}%
\providecommand \BibitemOpen [0]{}%
\providecommand \bibitemStop [0]{}%
\providecommand \bibitemNoStop [0]{.\EOS\space}%
\providecommand \EOS [0]{\spacefactor3000\relax}%
\providecommand \BibitemShut  [1]{\csname bibitem#1\endcsname}%
\let\auto@bib@innerbib\@empty
\bibitem [{\citenamefont {Sachdev}(2011)}]{sachdevQuantumPhaseTransitions2011}%
  \BibitemOpen
  \bibfield  {author} {\bibinfo {author} {\bibfnamefont {S.}~\bibnamefont {Sachdev}},\ }\href {https://doi.org/10.1017/CBO9780511973765} {\emph {\bibinfo {title} {Quantum {{Phase Transitions}}}}},\ \bibinfo {edition} {2nd}\ ed.\ (\bibinfo  {publisher} {Cambridge University Press},\ \bibinfo {year} {2011})\BibitemShut {NoStop}%
\bibitem [{\citenamefont {Senthil}()}]{senthil_book}%
  \BibitemOpen
  \bibfield  {author} {\bibinfo {author} {\bibfnamefont {T.}~\bibnamefont {Senthil}},\ }\bibinfo {title} {Deconfined quantum critical points: A review},\ in\ \href {https://doi.org/10.1142/9789811282386_0014} {\emph {\bibinfo {booktitle} {50 Years of the Renormalization Group}}},\ Chap.\ \bibinfo {chapter} {Chapter 14}, pp.\ \bibinfo {pages} {169--195}\BibitemShut {NoStop}%
\bibitem [{\citenamefont {Senthil}\ \emph {et~al.}(2005)\citenamefont {Senthil}, \citenamefont {Balents}, \citenamefont {Sachdev}, \citenamefont {Vishwanath},\ and\ \citenamefont {P.~A.~Fisher}}]{Senthil_2005}%
  \BibitemOpen
  \bibfield  {author} {\bibinfo {author} {\bibfnamefont {T.}~\bibnamefont {Senthil}}, \bibinfo {author} {\bibfnamefont {L.}~\bibnamefont {Balents}}, \bibinfo {author} {\bibfnamefont {S.}~\bibnamefont {Sachdev}}, \bibinfo {author} {\bibfnamefont {A.}~\bibnamefont {Vishwanath}},\ and\ \bibinfo {author} {\bibfnamefont {M.}~\bibnamefont {P.~A.~Fisher}},\ }\bibfield  {title} {\bibinfo {title} {Deconfined criticality critically defined},\ }\href {https://doi.org/10.1143/jpsjs.74s.1} {\bibfield  {journal} {\bibinfo  {journal} {Journal of the Physical Society of Japan}\ }\textbf {\bibinfo {volume} {74}},\ \bibinfo {pages} {1–9} (\bibinfo {year} {2005})}\BibitemShut {NoStop}%
\bibitem [{\citenamefont {Wang}\ \emph {et~al.}(2017)\citenamefont {Wang}, \citenamefont {Nahum}, \citenamefont {Metlitski}, \citenamefont {Xu},\ and\ \citenamefont {Senthil}}]{PhysRevX.7.031051}%
  \BibitemOpen
  \bibfield  {author} {\bibinfo {author} {\bibfnamefont {C.}~\bibnamefont {Wang}}, \bibinfo {author} {\bibfnamefont {A.}~\bibnamefont {Nahum}}, \bibinfo {author} {\bibfnamefont {M.~A.}\ \bibnamefont {Metlitski}}, \bibinfo {author} {\bibfnamefont {C.}~\bibnamefont {Xu}},\ and\ \bibinfo {author} {\bibfnamefont {T.}~\bibnamefont {Senthil}},\ }\bibfield  {title} {\bibinfo {title} {Deconfined quantum critical points: Symmetries and dualities},\ }\href {https://doi.org/10.1103/PhysRevX.7.031051} {\bibfield  {journal} {\bibinfo  {journal} {Phys. Rev. X}\ }\textbf {\bibinfo {volume} {7}},\ \bibinfo {pages} {031051} (\bibinfo {year} {2017})}\BibitemShut {NoStop}%
\bibitem [{\citenamefont {Bi}\ and\ \citenamefont {Senthil}(2019)}]{PhysRevX.9.021034}%
  \BibitemOpen
  \bibfield  {author} {\bibinfo {author} {\bibfnamefont {Z.}~\bibnamefont {Bi}}\ and\ \bibinfo {author} {\bibfnamefont {T.}~\bibnamefont {Senthil}},\ }\bibfield  {title} {\bibinfo {title} {Adventure in topological phase transitions in $3+1$-{D}: Non-abelian deconfined quantum criticalities and a possible duality},\ }\href {https://doi.org/10.1103/PhysRevX.9.021034} {\bibfield  {journal} {\bibinfo  {journal} {Phys. Rev. X}\ }\textbf {\bibinfo {volume} {9}},\ \bibinfo {pages} {021034} (\bibinfo {year} {2019})}\BibitemShut {NoStop}%
\bibitem [{\citenamefont {Senthil}\ \emph {et~al.}(2004{\natexlab{a}})\citenamefont {Senthil}, \citenamefont {Vishwanath}, \citenamefont {Balents}, \citenamefont {Sachdev},\ and\ \citenamefont {Fisher}}]{senthil_DQC}%
  \BibitemOpen
  \bibfield  {author} {\bibinfo {author} {\bibfnamefont {T.}~\bibnamefont {Senthil}}, \bibinfo {author} {\bibfnamefont {A.}~\bibnamefont {Vishwanath}}, \bibinfo {author} {\bibfnamefont {L.}~\bibnamefont {Balents}}, \bibinfo {author} {\bibfnamefont {S.}~\bibnamefont {Sachdev}},\ and\ \bibinfo {author} {\bibfnamefont {M.~P.~A.}\ \bibnamefont {Fisher}},\ }\bibfield  {title} {\bibinfo {title} {Deconfined quantum critical points},\ }\href {https://doi.org/10.1126/science.1091806} {\bibfield  {journal} {\bibinfo  {journal} {Science}\ }\textbf {\bibinfo {volume} {303}},\ \bibinfo {pages} {1490} (\bibinfo {year} {2004}{\natexlab{a}})}\BibitemShut {NoStop}%
\bibitem [{\citenamefont {Senthil}\ \emph {et~al.}(2004{\natexlab{b}})\citenamefont {Senthil}, \citenamefont {Balents}, \citenamefont {Sachdev}, \citenamefont {Vishwanath},\ and\ \citenamefont {Fisher}}]{PhysRevB.70.144407}%
  \BibitemOpen
  \bibfield  {author} {\bibinfo {author} {\bibfnamefont {T.}~\bibnamefont {Senthil}}, \bibinfo {author} {\bibfnamefont {L.}~\bibnamefont {Balents}}, \bibinfo {author} {\bibfnamefont {S.}~\bibnamefont {Sachdev}}, \bibinfo {author} {\bibfnamefont {A.}~\bibnamefont {Vishwanath}},\ and\ \bibinfo {author} {\bibfnamefont {M.~P.~A.}\ \bibnamefont {Fisher}},\ }\bibfield  {title} {\bibinfo {title} {Quantum criticality beyond the landau-ginzburg-wilson paradigm},\ }\href {https://doi.org/10.1103/PhysRevB.70.144407} {\bibfield  {journal} {\bibinfo  {journal} {Phys. Rev. B}\ }\textbf {\bibinfo {volume} {70}},\ \bibinfo {pages} {144407} (\bibinfo {year} {2004}{\natexlab{b}})}\BibitemShut {NoStop}%
\bibitem [{\citenamefont {Sandvik}(2007)}]{PhysRevLett.98.227202}%
  \BibitemOpen
  \bibfield  {author} {\bibinfo {author} {\bibfnamefont {A.~W.}\ \bibnamefont {Sandvik}},\ }\bibfield  {title} {\bibinfo {title} {Evidence for deconfined quantum criticality in a two-dimensional heisenberg model with four-spin interactions},\ }\href {https://doi.org/10.1103/PhysRevLett.98.227202} {\bibfield  {journal} {\bibinfo  {journal} {Phys. Rev. Lett.}\ }\textbf {\bibinfo {volume} {98}},\ \bibinfo {pages} {227202} (\bibinfo {year} {2007})}\BibitemShut {NoStop}%
\bibitem [{\citenamefont {Nahum}\ \emph {et~al.}(2015)\citenamefont {Nahum}, \citenamefont {Chalker}, \citenamefont {Serna}, \citenamefont {Ortu\~no},\ and\ \citenamefont {Somoza}}]{PhysRevX.5.041048}%
  \BibitemOpen
  \bibfield  {author} {\bibinfo {author} {\bibfnamefont {A.}~\bibnamefont {Nahum}}, \bibinfo {author} {\bibfnamefont {J.~T.}\ \bibnamefont {Chalker}}, \bibinfo {author} {\bibfnamefont {P.}~\bibnamefont {Serna}}, \bibinfo {author} {\bibfnamefont {M.}~\bibnamefont {Ortu\~no}},\ and\ \bibinfo {author} {\bibfnamefont {A.~M.}\ \bibnamefont {Somoza}},\ }\bibfield  {title} {\bibinfo {title} {Deconfined quantum criticality, scaling violations, and classical loop models},\ }\href {https://doi.org/10.1103/PhysRevX.5.041048} {\bibfield  {journal} {\bibinfo  {journal} {Phys. Rev. X}\ }\textbf {\bibinfo {volume} {5}},\ \bibinfo {pages} {041048} (\bibinfo {year} {2015})}\BibitemShut {NoStop}%
\bibitem [{\citenamefont {Sandvik}\ and\ \citenamefont {Zhao}(2020)}]{Sandvik_2020}%
  \BibitemOpen
  \bibfield  {author} {\bibinfo {author} {\bibfnamefont {A.~W.}\ \bibnamefont {Sandvik}}\ and\ \bibinfo {author} {\bibfnamefont {B.}~\bibnamefont {Zhao}},\ }\bibfield  {title} {\bibinfo {title} {Consistent scaling exponents at the deconfined quantum-critical point*},\ }\href {https://doi.org/10.1088/0256-307x/37/5/057502} {\bibfield  {journal} {\bibinfo  {journal} {Chinese Physics Letters}\ }\textbf {\bibinfo {volume} {37}},\ \bibinfo {pages} {057502} (\bibinfo {year} {2020})}\BibitemShut {NoStop}%
\bibitem [{\citenamefont {D'Emidio}\ and\ \citenamefont {Sandvik}(2024)}]{PhysRevLett.133.166702}%
  \BibitemOpen
  \bibfield  {author} {\bibinfo {author} {\bibfnamefont {J.}~\bibnamefont {D'Emidio}}\ and\ \bibinfo {author} {\bibfnamefont {A.~W.}\ \bibnamefont {Sandvik}},\ }\bibfield  {title} {\bibinfo {title} {Entanglement entropy and deconfined criticality: Emergent so(5) symmetry and proper lattice bipartition},\ }\href {https://doi.org/10.1103/PhysRevLett.133.166702} {\bibfield  {journal} {\bibinfo  {journal} {Phys. Rev. Lett.}\ }\textbf {\bibinfo {volume} {133}},\ \bibinfo {pages} {166702} (\bibinfo {year} {2024})}\BibitemShut {NoStop}%
\bibitem [{\citenamefont {Baxter}(1972{\natexlab{a}})}]{BAXTER1972323}%
  \BibitemOpen
  \bibfield  {author} {\bibinfo {author} {\bibfnamefont {R.~J.}\ \bibnamefont {Baxter}},\ }\bibfield  {title} {\bibinfo {title} {One-dimensional anisotropic heisenberg chain},\ }\href {https://doi.org/https://doi.org/10.1016/0003-4916(72)90270-9} {\bibfield  {journal} {\bibinfo  {journal} {Annals of Physics}\ }\textbf {\bibinfo {volume} {70}},\ \bibinfo {pages} {323} (\bibinfo {year} {1972}{\natexlab{a}})}\BibitemShut {NoStop}%
\bibitem [{\citenamefont {Baxter}(1972{\natexlab{b}})}]{BAXTER1972193}%
  \BibitemOpen
  \bibfield  {author} {\bibinfo {author} {\bibfnamefont {R.~J.}\ \bibnamefont {Baxter}},\ }\bibfield  {title} {\bibinfo {title} {Partition function of the eight-vertex lattice model},\ }\href {https://doi.org/https://doi.org/10.1016/0003-4916(72)90335-1} {\bibfield  {journal} {\bibinfo  {journal} {Annals of Physics}\ }\textbf {\bibinfo {volume} {70}},\ \bibinfo {pages} {193} (\bibinfo {year} {1972}{\natexlab{b}})}\BibitemShut {NoStop}%
\bibitem [{\citenamefont {Chepiga}\ and\ \citenamefont {Mila}(2023)}]{8vertex}%
  \BibitemOpen
  \bibfield  {author} {\bibinfo {author} {\bibfnamefont {N.}~\bibnamefont {Chepiga}}\ and\ \bibinfo {author} {\bibfnamefont {F.}~\bibnamefont {Mila}},\ }\bibfield  {title} {\bibinfo {title} {Eight-vertex criticality in the interacting kitaev chain},\ }\href {https://doi.org/10.1103/PhysRevB.107.L081106} {\bibfield  {journal} {\bibinfo  {journal} {Phys. Rev. B}\ }\textbf {\bibinfo {volume} {107}},\ \bibinfo {pages} {L081106} (\bibinfo {year} {2023})}\BibitemShut {NoStop}%
\bibitem [{\citenamefont {Michaud}\ \emph {et~al.}(2012)\citenamefont {Michaud}, \citenamefont {Vernay}, \citenamefont {Manmana},\ and\ \citenamefont {Mila}}]{michaudAntiferromagneticSpinSChains2012}%
  \BibitemOpen
  \bibfield  {author} {\bibinfo {author} {\bibfnamefont {F.}~\bibnamefont {Michaud}}, \bibinfo {author} {\bibfnamefont {F.}~\bibnamefont {Vernay}}, \bibinfo {author} {\bibfnamefont {S.~R.}\ \bibnamefont {Manmana}},\ and\ \bibinfo {author} {\bibfnamefont {F.}~\bibnamefont {Mila}},\ }\bibfield  {title} {\bibinfo {title} {Antiferromagnetic spin-{$S$} chains with exactly dimerized ground states},\ }\href {https://doi.org/10.1103/PhysRevLett.108.127202} {\bibfield  {journal} {\bibinfo  {journal} {Physical Review Letters}\ }\textbf {\bibinfo {volume} {108}},\ \bibinfo {pages} {127202} (\bibinfo {year} {2012})},\ \Eprint {https://arxiv.org/abs/1110.3394} {1110.3394 [cond-mat]} \BibitemShut {NoStop}%
\bibitem [{\citenamefont {Majumdar}\ and\ \citenamefont {Ghosh}(1969)}]{majumdarNextNearestNeighborInteractionLinear1969}%
  \BibitemOpen
  \bibfield  {author} {\bibinfo {author} {\bibfnamefont {C.~K.}\ \bibnamefont {Majumdar}}\ and\ \bibinfo {author} {\bibfnamefont {D.~K.}\ \bibnamefont {Ghosh}},\ }\bibfield  {title} {\bibinfo {title} {On {{Next-Nearest-Neighbor Interaction}} in {{Linear Chain}}. {{I}}},\ }\href {https://doi.org/10.1063/1.1664978} {\bibfield  {journal} {\bibinfo  {journal} {Journal of Mathematical Physics}\ }\textbf {\bibinfo {volume} {10}},\ \bibinfo {pages} {1388} (\bibinfo {year} {1969})}\BibitemShut {NoStop}%
\bibitem [{\citenamefont {Haldane}(1983)}]{HALDANE1983464}%
  \BibitemOpen
  \bibfield  {author} {\bibinfo {author} {\bibfnamefont {F.}~\bibnamefont {Haldane}},\ }\bibfield  {title} {\bibinfo {title} {Continuum dynamics of the 1-d heisenberg antiferromagnet: Identification with the o(3) nonlinear sigma model},\ }\href {https://doi.org/https://doi.org/10.1016/0375-9601(83)90631-X} {\bibfield  {journal} {\bibinfo  {journal} {Physics Letters A}\ }\textbf {\bibinfo {volume} {93}},\ \bibinfo {pages} {464} (\bibinfo {year} {1983})}\BibitemShut {NoStop}%
\bibitem [{\citenamefont {Kennedy}(1990)}]{TKennedy_1990}%
  \BibitemOpen
  \bibfield  {author} {\bibinfo {author} {\bibfnamefont {T.}~\bibnamefont {Kennedy}},\ }\bibfield  {title} {\bibinfo {title} {Exact diagonalisations of open spin-1 chains},\ }\href {https://doi.org/10.1088/0953-8984/2/26/010} {\bibfield  {journal} {\bibinfo  {journal} {Journal of Physics: Condensed Matter}\ }\textbf {\bibinfo {volume} {2}},\ \bibinfo {pages} {5737} (\bibinfo {year} {1990})}\BibitemShut {NoStop}%
\bibitem [{\citenamefont {Hagiwara}\ \emph {et~al.}(1990)\citenamefont {Hagiwara}, \citenamefont {Katsumata}, \citenamefont {Affleck}, \citenamefont {Halperin},\ and\ \citenamefont {Renard}}]{PhysRevLett.65.3181}%
  \BibitemOpen
  \bibfield  {author} {\bibinfo {author} {\bibfnamefont {M.}~\bibnamefont {Hagiwara}}, \bibinfo {author} {\bibfnamefont {K.}~\bibnamefont {Katsumata}}, \bibinfo {author} {\bibfnamefont {I.}~\bibnamefont {Affleck}}, \bibinfo {author} {\bibfnamefont {B.~I.}\ \bibnamefont {Halperin}},\ and\ \bibinfo {author} {\bibfnamefont {J.~P.}\ \bibnamefont {Renard}},\ }\bibfield  {title} {\bibinfo {title} {Observation of {$S=1/2$} degrees of freedom in an {$S=1$} linear-chain heisenberg antiferromagnet},\ }\href {https://doi.org/10.1103/PhysRevLett.65.3181} {\bibfield  {journal} {\bibinfo  {journal} {Phys. Rev. Lett.}\ }\textbf {\bibinfo {volume} {65}},\ \bibinfo {pages} {3181} (\bibinfo {year} {1990})}\BibitemShut {NoStop}%
\bibitem [{\citenamefont {Affleck}\ and\ \citenamefont {Haldane}(1987)}]{affleckCriticalTheoryQuantum1987b}%
  \BibitemOpen
  \bibfield  {author} {\bibinfo {author} {\bibfnamefont {I.}~\bibnamefont {Affleck}}\ and\ \bibinfo {author} {\bibfnamefont {F.~D.~M.}\ \bibnamefont {Haldane}},\ }\bibfield  {title} {\bibinfo {title} {Critical theory of quantum spin chains},\ }\href {https://doi.org/10.1103/PhysRevB.36.5291} {\bibfield  {journal} {\bibinfo  {journal} {Physical Review B}\ }\textbf {\bibinfo {volume} {36}},\ \bibinfo {pages} {5291} (\bibinfo {year} {1987})}\BibitemShut {NoStop}%
\bibitem [{\citenamefont {Babujian}(1982)}]{babujian_1982}%
  \BibitemOpen
  \bibfield  {author} {\bibinfo {author} {\bibfnamefont {H.~M.}\ \bibnamefont {Babujian}},\ }\bibfield  {title} {\bibinfo {title} {Exact solution of the one-dimensional isotropic {{Heisenberg}} chain with arbitrary spins {{{\emph{S}}}}},\ }\href {https://doi.org/10.1016/0375-9601(82)90403-0} {\bibfield  {journal} {\bibinfo  {journal} {Physics Letters A}\ }\textbf {\bibinfo {volume} {90}},\ \bibinfo {pages} {479} (\bibinfo {year} {1982})}\BibitemShut {NoStop}%
\bibitem [{\citenamefont {Takhtajan}(1982)}]{takhtajan_1982}%
  \BibitemOpen
  \bibfield  {author} {\bibinfo {author} {\bibfnamefont {L.~A.}\ \bibnamefont {Takhtajan}},\ }\bibfield  {title} {\bibinfo {title} {The picture of low-lying excitations in the isotropic {{Heisenberg}} chain of arbitrary spins},\ }\href {https://doi.org/10.1016/0375-9601(82)90764-2} {\bibfield  {journal} {\bibinfo  {journal} {Physics Letters A}\ }\textbf {\bibinfo {volume} {87}},\ \bibinfo {pages} {479} (\bibinfo {year} {1982})}\BibitemShut {NoStop}%
\bibitem [{\citenamefont {Botet}\ \emph {et~al.}(1983)\citenamefont {Botet}, \citenamefont {Jullien},\ and\ \citenamefont {Kolb}}]{PhysRevB.28.3914}%
  \BibitemOpen
  \bibfield  {author} {\bibinfo {author} {\bibfnamefont {R.}~\bibnamefont {Botet}}, \bibinfo {author} {\bibfnamefont {R.}~\bibnamefont {Jullien}},\ and\ \bibinfo {author} {\bibfnamefont {M.}~\bibnamefont {Kolb}},\ }\bibfield  {title} {\bibinfo {title} {Finite-size-scaling study of the spin-1 heisenberg-ising chain with uniaxial anisotropy},\ }\href {https://doi.org/10.1103/PhysRevB.28.3914} {\bibfield  {journal} {\bibinfo  {journal} {Phys. Rev. B}\ }\textbf {\bibinfo {volume} {28}},\ \bibinfo {pages} {3914} (\bibinfo {year} {1983})}\BibitemShut {NoStop}%
\bibitem [{\citenamefont {Chen}\ \emph {et~al.}(2003)\citenamefont {Chen}, \citenamefont {Hida},\ and\ \citenamefont {Sanctuary}}]{PhysRevB.67.104401}%
  \BibitemOpen
  \bibfield  {author} {\bibinfo {author} {\bibfnamefont {W.}~\bibnamefont {Chen}}, \bibinfo {author} {\bibfnamefont {K.}~\bibnamefont {Hida}},\ and\ \bibinfo {author} {\bibfnamefont {B.~C.}\ \bibnamefont {Sanctuary}},\ }\bibfield  {title} {\bibinfo {title} {Ground-state phase diagram of {$S=1$} $\mathrm{XXZ}$ chains with uniaxial single-ion-type anisotropy},\ }\href {https://doi.org/10.1103/PhysRevB.67.104401} {\bibfield  {journal} {\bibinfo  {journal} {Phys. Rev. B}\ }\textbf {\bibinfo {volume} {67}},\ \bibinfo {pages} {104401} (\bibinfo {year} {2003})}\BibitemShut {NoStop}%
\bibitem [{\citenamefont {Schulz}(1986)}]{PhysRevB.34.6372}%
  \BibitemOpen
  \bibfield  {author} {\bibinfo {author} {\bibfnamefont {H.~J.}\ \bibnamefont {Schulz}},\ }\bibfield  {title} {\bibinfo {title} {Phase diagrams and correlation exponents for quantum spin chains of arbitrary spin quantum number},\ }\href {https://doi.org/10.1103/PhysRevB.34.6372} {\bibfield  {journal} {\bibinfo  {journal} {Phys. Rev. B}\ }\textbf {\bibinfo {volume} {34}},\ \bibinfo {pages} {6372} (\bibinfo {year} {1986})}\BibitemShut {NoStop}%
\bibitem [{\citenamefont {Ver\'{\i}ssimo}\ \emph {et~al.}(2021)\citenamefont {Ver\'{\i}ssimo}, \citenamefont {Pereira},\ and\ \citenamefont {Lyra}}]{PhysRevB.104.024409}%
  \BibitemOpen
  \bibfield  {author} {\bibinfo {author} {\bibfnamefont {L.~M.}\ \bibnamefont {Ver\'{\i}ssimo}}, \bibinfo {author} {\bibfnamefont {M.~S.~S.}\ \bibnamefont {Pereira}},\ and\ \bibinfo {author} {\bibfnamefont {M.~L.}\ \bibnamefont {Lyra}},\ }\bibfield  {title} {\bibinfo {title} {Tangential finite-size scaling at the gaussian topological transition in the quantum spin-1 anisotropic chain},\ }\href {https://doi.org/10.1103/PhysRevB.104.024409} {\bibfield  {journal} {\bibinfo  {journal} {Phys. Rev. B}\ }\textbf {\bibinfo {volume} {104}},\ \bibinfo {pages} {024409} (\bibinfo {year} {2021})}\BibitemShut {NoStop}%
\bibitem [{\citenamefont {De~Chiara}\ \emph {et~al.}(2011)\citenamefont {De~Chiara}, \citenamefont {Lewenstein},\ and\ \citenamefont {Sanpera}}]{PhysRevB.84.054451}%
  \BibitemOpen
  \bibfield  {author} {\bibinfo {author} {\bibfnamefont {G.}~\bibnamefont {De~Chiara}}, \bibinfo {author} {\bibfnamefont {M.}~\bibnamefont {Lewenstein}},\ and\ \bibinfo {author} {\bibfnamefont {A.}~\bibnamefont {Sanpera}},\ }\bibfield  {title} {\bibinfo {title} {Bilinear-biquadratic spin-1 chain undergoing quadratic zeeman effect},\ }\href {https://doi.org/10.1103/PhysRevB.84.054451} {\bibfield  {journal} {\bibinfo  {journal} {Phys. Rev. B}\ }\textbf {\bibinfo {volume} {84}},\ \bibinfo {pages} {054451} (\bibinfo {year} {2011})}\BibitemShut {NoStop}%
\bibitem [{\citenamefont {Degli Esposti~Boschi}\ \emph {et~al.}(2003)\citenamefont {Degli Esposti~Boschi}, \citenamefont {Ercolessi}, \citenamefont {Ortolani},\ and\ \citenamefont {Roncaglia}}]{ercolessi}%
  \BibitemOpen
  \bibfield  {author} {\bibinfo {author} {\bibfnamefont {C.}~\bibnamefont {Degli Esposti~Boschi}}, \bibinfo {author} {\bibfnamefont {E.}~\bibnamefont {Ercolessi}}, \bibinfo {author} {\bibfnamefont {F.}~\bibnamefont {Ortolani}},\ and\ \bibinfo {author} {\bibfnamefont {M.}~\bibnamefont {Roncaglia}},\ }\bibfield  {title} {\bibinfo {title} {On $\mathsf{c = 1}$ critical phases in anisotropic spin-1 chains},\ }\href {https://doi.org/10.1140/epjb/e2003-00299-7} {\bibfield  {journal} {\bibinfo  {journal} {The European Physical Journal B - Condensed Matter}\ }\textbf {\bibinfo {volume} {35}},\ \bibinfo {pages} {465–473} (\bibinfo {year} {2003})}\BibitemShut {NoStop}%
\bibitem [{\citenamefont {Hu}\ \emph {et~al.}(2011)\citenamefont {Hu}, \citenamefont {Normand}, \citenamefont {Wang},\ and\ \citenamefont {Yu}}]{PhysRevB.84.220402}%
  \BibitemOpen
  \bibfield  {author} {\bibinfo {author} {\bibfnamefont {S.}~\bibnamefont {Hu}}, \bibinfo {author} {\bibfnamefont {B.}~\bibnamefont {Normand}}, \bibinfo {author} {\bibfnamefont {X.}~\bibnamefont {Wang}},\ and\ \bibinfo {author} {\bibfnamefont {L.}~\bibnamefont {Yu}},\ }\bibfield  {title} {\bibinfo {title} {Accurate determination of the gaussian transition in spin-1 chains with single-ion anisotropy},\ }\href {https://doi.org/10.1103/PhysRevB.84.220402} {\bibfield  {journal} {\bibinfo  {journal} {Phys. Rev. B}\ }\textbf {\bibinfo {volume} {84}},\ \bibinfo {pages} {220402} (\bibinfo {year} {2011})}\BibitemShut {NoStop}%
\bibitem [{\citenamefont {Haldane}(1982)}]{haldaneSpontaneousDimerization$Sfrac12$1982}%
  \BibitemOpen
  \bibfield  {author} {\bibinfo {author} {\bibfnamefont {F.~D.~M.}\ \bibnamefont {Haldane}},\ }\bibfield  {title} {\bibinfo {title} {Spontaneous dimerization in the {$S=\frac{1}{2}$} {{Heisenberg}} antiferromagnetic chain with competing interactions},\ }\href {https://doi.org/10.1103/PhysRevB.25.4925} {\bibfield  {journal} {\bibinfo  {journal} {Phys. Rev. B}\ }\textbf {\bibinfo {volume} {25}},\ \bibinfo {pages} {4925} (\bibinfo {year} {1982})}\BibitemShut {NoStop}%
\bibitem [{\citenamefont {Nomura}\ and\ \citenamefont {Okamoto}(1994)}]{nomuraCriticalProperties11994}%
  \BibitemOpen
  \bibfield  {author} {\bibinfo {author} {\bibfnamefont {K.}~\bibnamefont {Nomura}}\ and\ \bibinfo {author} {\bibfnamefont {K.}~\bibnamefont {Okamoto}},\ }\bibfield  {title} {\bibinfo {title} {Critical properties of {$S= 1/2$} antiferromagnetic {{XXZ}} chain with next-nearest-neighbour interactions},\ }\href {https://doi.org/10.1088/0305-4470/27/17/012} {\bibfield  {journal} {\bibinfo  {journal} {J. Phys. A: Math. Gen.}\ }\textbf {\bibinfo {volume} {27}},\ \bibinfo {pages} {5773} (\bibinfo {year} {1994})}\BibitemShut {NoStop}%
\bibitem [{SM()}]{SM}%
  \BibitemOpen
  \href@noop {} {}\bibinfo {note} {See Supplemental Material, which includes Refs.~\cite{ercolessi,PhysRevB.84.220402,scan_DMRG,PhysRevB.96.054425}, for technical details on the DMRG and scan-DMRG algorithms used in the paper, the numerical results supporting the Ising transition between the Haldane and Ising-AFM phases, the topological nature of the Ising-AFM, the Gaussian transition, and the details in the protocol used to extract the correlation length $\xi$.}\BibitemShut {Stop}%
\bibitem [{\citenamefont {White}(1992)}]{whiteDensityMatrixFormulation1992a}%
  \BibitemOpen
  \bibfield  {author} {\bibinfo {author} {\bibfnamefont {S.~R.}\ \bibnamefont {White}},\ }\bibfield  {title} {\bibinfo {title} {Density matrix formulation for quantum renormalization groups},\ }\href {https://doi.org/10.1103/PhysRevLett.69.2863} {\bibfield  {journal} {\bibinfo  {journal} {Physical Review Letters}\ }\textbf {\bibinfo {volume} {69}},\ \bibinfo {pages} {2863} (\bibinfo {year} {1992})}\BibitemShut {NoStop}%
\bibitem [{\citenamefont {White}(1993)}]{whiteDensitymatrixAlgorithmsQuantum1993a}%
  \BibitemOpen
  \bibfield  {author} {\bibinfo {author} {\bibfnamefont {S.~R.}\ \bibnamefont {White}},\ }\bibfield  {title} {\bibinfo {title} {Density-matrix algorithms for quantum renormalization groups},\ }\href {https://doi.org/10.1103/PhysRevB.48.10345} {\bibfield  {journal} {\bibinfo  {journal} {Physical Review B}\ }\textbf {\bibinfo {volume} {48}},\ \bibinfo {pages} {10345} (\bibinfo {year} {1993})}\BibitemShut {NoStop}%
\bibitem [{\citenamefont {Zamolodchikov}\ and\ \citenamefont {Fateev}(1985)}]{zamolodchikovNonlocalParafermionCurrents1985b}%
  \BibitemOpen
  \bibfield  {author} {\bibinfo {author} {\bibfnamefont {A.~B.}\ \bibnamefont {Zamolodchikov}}\ and\ \bibinfo {author} {\bibfnamefont {V.~A.}\ \bibnamefont {Fateev}},\ }\bibfield  {title} {\bibinfo {title} {Nonlocal (parafermion) currents in two-dimensional conformal quantum field theory and self-dual critical points in ${Z_N}$-symmetric statistical systems},\ }\href@noop {} {\bibfield  {journal} {\bibinfo  {journal} {Sov. Phys. - JETP (Engl. Transl.); (United States)}\ }\textbf {\bibinfo {volume} {62:2}} (\bibinfo {year} {1985})}\BibitemShut {NoStop}%
\bibitem [{\citenamefont {Chepiga}\ \emph {et~al.}(2016{\natexlab{a}})\citenamefont {Chepiga}, \citenamefont {Affleck},\ and\ \citenamefont {Mila}}]{chepigaDimerizationTransitionsSpin12016a}%
  \BibitemOpen
  \bibfield  {author} {\bibinfo {author} {\bibfnamefont {N.}~\bibnamefont {Chepiga}}, \bibinfo {author} {\bibfnamefont {I.}~\bibnamefont {Affleck}},\ and\ \bibinfo {author} {\bibfnamefont {F.}~\bibnamefont {Mila}},\ }\bibfield  {title} {\bibinfo {title} {Dimerization transitions in spin-1 chains},\ }\href {https://doi.org/10.1103/PhysRevB.93.241108} {\bibfield  {journal} {\bibinfo  {journal} {Physical Review B}\ }\textbf {\bibinfo {volume} {93}},\ \bibinfo {pages} {241108} (\bibinfo {year} {2016}{\natexlab{a}})}\BibitemShut {NoStop}%
\bibitem [{\citenamefont {Chepiga}\ \emph {et~al.}(2016{\natexlab{b}})\citenamefont {Chepiga}, \citenamefont {Affleck},\ and\ \citenamefont {Mila}}]{chepigaCommentFrustrationMulticriticality2016a}%
  \BibitemOpen
  \bibfield  {author} {\bibinfo {author} {\bibfnamefont {N.}~\bibnamefont {Chepiga}}, \bibinfo {author} {\bibfnamefont {I.}~\bibnamefont {Affleck}},\ and\ \bibinfo {author} {\bibfnamefont {F.}~\bibnamefont {Mila}},\ }\bibfield  {title} {\bibinfo {title} {Comment on ``{{Frustration}} and multicriticality in the antiferromagnetic spin-1 chain''},\ }\href {https://doi.org/10.1103/PhysRevB.94.136401} {\bibfield  {journal} {\bibinfo  {journal} {Physical Review B}\ }\textbf {\bibinfo {volume} {94}},\ \bibinfo {pages} {136401} (\bibinfo {year} {2016}{\natexlab{b}})}\BibitemShut {NoStop}%
\bibitem [{\citenamefont {Laflorencie}\ \emph {et~al.}(2006)\citenamefont {Laflorencie}, \citenamefont {S\o{}rensen}, \citenamefont {Chang},\ and\ \citenamefont {Affleck}}]{PhysRevLett.96.100603}%
  \BibitemOpen
  \bibfield  {author} {\bibinfo {author} {\bibfnamefont {N.}~\bibnamefont {Laflorencie}}, \bibinfo {author} {\bibfnamefont {E.~S.}\ \bibnamefont {S\o{}rensen}}, \bibinfo {author} {\bibfnamefont {M.-S.}\ \bibnamefont {Chang}},\ and\ \bibinfo {author} {\bibfnamefont {I.}~\bibnamefont {Affleck}},\ }\bibfield  {title} {\bibinfo {title} {Boundary effects in the critical scaling of entanglement entropy in {1D} systems},\ }\href {https://doi.org/10.1103/PhysRevLett.96.100603} {\bibfield  {journal} {\bibinfo  {journal} {Phys. Rev. Lett.}\ }\textbf {\bibinfo {volume} {96}},\ \bibinfo {pages} {100603} (\bibinfo {year} {2006})}\BibitemShut {NoStop}%
\bibitem [{\citenamefont {Capponi}\ \emph {et~al.}(2013)\citenamefont {Capponi}, \citenamefont {Lecheminant},\ and\ \citenamefont {Moliner}}]{capponiQuantumPhaseTransitions2013}%
  \BibitemOpen
  \bibfield  {author} {\bibinfo {author} {\bibfnamefont {S.}~\bibnamefont {Capponi}}, \bibinfo {author} {\bibfnamefont {P.}~\bibnamefont {Lecheminant}},\ and\ \bibinfo {author} {\bibfnamefont {M.}~\bibnamefont {Moliner}},\ }\bibfield  {title} {\bibinfo {title} {Quantum phase transitions in multileg spin ladders with ring exchange},\ }\href {https://doi.org/10.1103/PhysRevB.88.075132} {\bibfield  {journal} {\bibinfo  {journal} {Physical Review B}\ }\textbf {\bibinfo {volume} {88}},\ \bibinfo {pages} {075132} (\bibinfo {year} {2013})}\BibitemShut {NoStop}%
\bibitem [{\citenamefont {Calabrese}\ and\ \citenamefont {Cardy}(2004)}]{calabreseEntanglementEntropyQuantum2004}%
  \BibitemOpen
  \bibfield  {author} {\bibinfo {author} {\bibfnamefont {P.}~\bibnamefont {Calabrese}}\ and\ \bibinfo {author} {\bibfnamefont {J.}~\bibnamefont {Cardy}},\ }\bibfield  {title} {\bibinfo {title} {Entanglement {{Entropy}} and {{Quantum Field Theory}}},\ }\href {https://doi.org/10.1088/1742-5468/2004/06/P06002} {\bibfield  {journal} {\bibinfo  {journal} {Journal of Statistical Mechanics: Theory and Experiment}\ }\textbf {\bibinfo {volume} {2004}},\ \bibinfo {pages} {P06002} (\bibinfo {year} {2004})}\BibitemShut {NoStop}%
\bibitem [{\citenamefont {Chepiga}(2024)}]{scan_DMRG}%
  \BibitemOpen
  \bibfield  {author} {\bibinfo {author} {\bibfnamefont {N.}~\bibnamefont {Chepiga}},\ }\bibfield  {title} {\bibinfo {title} {Probing universal critical scaling with scan density matrix renormalization group},\ }\href {https://doi.org/10.1103/PhysRevB.110.144401} {\bibfield  {journal} {\bibinfo  {journal} {Phys. Rev. B}\ }\textbf {\bibinfo {volume} {110}},\ \bibinfo {pages} {144401} (\bibinfo {year} {2024})}\BibitemShut {NoStop}%
\bibitem [{\citenamefont {Di~Francesco}\ \emph {et~al.}(1997)\citenamefont {Di~Francesco}, \citenamefont {Mathieu},\ and\ \citenamefont {S{\'e}n{\'e}chal}}]{difrancescoConformalFieldTheory1997}%
  \BibitemOpen
  \bibfield  {author} {\bibinfo {author} {\bibfnamefont {P.}~\bibnamefont {Di~Francesco}}, \bibinfo {author} {\bibfnamefont {P.}~\bibnamefont {Mathieu}},\ and\ \bibinfo {author} {\bibfnamefont {D.}~\bibnamefont {S{\'e}n{\'e}chal}},\ }\href {https://doi.org/10.1007/978-1-4612-2256-9} {\emph {\bibinfo {title} {Conformal Field Theory}}},\ Graduate {{Texts}} in {{Contemporary Physics}}\ (\bibinfo  {publisher} {Springer},\ \bibinfo {year} {1997})\BibitemShut {NoStop}%
\bibitem [{\citenamefont {Kohmoto}\ \emph {et~al.}(1981)\citenamefont {Kohmoto}, \citenamefont {den Nijs},\ and\ \citenamefont {Kadanoff}}]{PhysRevB.24.5229}%
  \BibitemOpen
  \bibfield  {author} {\bibinfo {author} {\bibfnamefont {M.}~\bibnamefont {Kohmoto}}, \bibinfo {author} {\bibfnamefont {M.}~\bibnamefont {den Nijs}},\ and\ \bibinfo {author} {\bibfnamefont {L.~P.}\ \bibnamefont {Kadanoff}},\ }\bibfield  {title} {\bibinfo {title} {Hamiltonian studies of the $d=2$ {Ashkin-Teller} model},\ }\href {https://doi.org/10.1103/PhysRevB.24.5229} {\bibfield  {journal} {\bibinfo  {journal} {Phys. Rev. B}\ }\textbf {\bibinfo {volume} {24}},\ \bibinfo {pages} {5229} (\bibinfo {year} {1981})}\BibitemShut {NoStop}%
\bibitem [{\citenamefont {Bridgeman}\ \emph {et~al.}(2015)\citenamefont {Bridgeman}, \citenamefont {O'Brien}, \citenamefont {Bartlett},\ and\ \citenamefont {Doherty}}]{PhysRevB.91.165129}%
  \BibitemOpen
  \bibfield  {author} {\bibinfo {author} {\bibfnamefont {J.~C.}\ \bibnamefont {Bridgeman}}, \bibinfo {author} {\bibfnamefont {A.}~\bibnamefont {O'Brien}}, \bibinfo {author} {\bibfnamefont {S.~D.}\ \bibnamefont {Bartlett}},\ and\ \bibinfo {author} {\bibfnamefont {A.~C.}\ \bibnamefont {Doherty}},\ }\bibfield  {title} {\bibinfo {title} {Multiscale entanglement renormalization ansatz for spin chains with continuously varying criticality},\ }\href {https://doi.org/10.1103/PhysRevB.91.165129} {\bibfield  {journal} {\bibinfo  {journal} {Phys. Rev. B}\ }\textbf {\bibinfo {volume} {91}},\ \bibinfo {pages} {165129} (\bibinfo {year} {2015})}\BibitemShut {NoStop}%
\bibitem [{\citenamefont {den Nijs}(1988)}]{Den_Nijs}%
  \BibitemOpen
  \bibfield  {author} {\bibinfo {author} {\bibfnamefont {M.}~\bibnamefont {den Nijs}},\ }\bibfield  {title} {\bibinfo {title} {The domain wall theory of two-dimensional commensurate-incommensurate phase transitions},\ }\href@noop {} {\bibfield  {journal} {\bibinfo  {journal} {Phase Transitions and Critical Phenomena}\ }\textbf {\bibinfo {volume} {12}},\ \bibinfo {pages} {219} (\bibinfo {year} {1988})}\BibitemShut {NoStop}%
\bibitem [{\citenamefont {Chepiga}\ and\ \citenamefont {Laflorencie}(2023)}]{10.21468/SciPostPhys.14.6.152}%
  \BibitemOpen
  \bibfield  {author} {\bibinfo {author} {\bibfnamefont {N.}~\bibnamefont {Chepiga}}\ and\ \bibinfo {author} {\bibfnamefont {N.}~\bibnamefont {Laflorencie}},\ }\bibfield  {title} {\bibinfo {title} {{Topological and quantum critical properties of the interacting Majorana chain model}},\ }\href {https://doi.org/10.21468/SciPostPhys.14.6.152} {\bibfield  {journal} {\bibinfo  {journal} {SciPost Phys.}\ }\textbf {\bibinfo {volume} {14}},\ \bibinfo {pages} {152} (\bibinfo {year} {2023})}\BibitemShut {NoStop}%
\bibitem [{\citenamefont {Chepiga}(2023)}]{PhysRevB.108.054509}%
  \BibitemOpen
  \bibfield  {author} {\bibinfo {author} {\bibfnamefont {N.}~\bibnamefont {Chepiga}},\ }\bibfield  {title} {\bibinfo {title} {Critical properties of the majorana chain with competing interactions},\ }\href {https://doi.org/10.1103/PhysRevB.108.054509} {\bibfield  {journal} {\bibinfo  {journal} {Phys. Rev. B}\ }\textbf {\bibinfo {volume} {108}},\ \bibinfo {pages} {054509} (\bibinfo {year} {2023})}\BibitemShut {NoStop}%
\bibitem [{\citenamefont {Ogino}\ \emph {et~al.}(2021{\natexlab{a}})\citenamefont {Ogino}, \citenamefont {Kaneko}, \citenamefont {Morita}, \citenamefont {Furukawa},\ and\ \citenamefont {Kawashima}}]{oginoContinuousPhaseTransition2021}%
  \BibitemOpen
  \bibfield  {author} {\bibinfo {author} {\bibfnamefont {T.}~\bibnamefont {Ogino}}, \bibinfo {author} {\bibfnamefont {R.}~\bibnamefont {Kaneko}}, \bibinfo {author} {\bibfnamefont {S.}~\bibnamefont {Morita}}, \bibinfo {author} {\bibfnamefont {S.}~\bibnamefont {Furukawa}},\ and\ \bibinfo {author} {\bibfnamefont {N.}~\bibnamefont {Kawashima}},\ }\bibfield  {title} {\bibinfo {title} {Continuous phase transition between {{N\'eel}} and valence bond solid phases in a {{J-Q-like}} spin ladder system},\ }\href {https://doi.org/10.1103/PhysRevB.103.085117} {\bibfield  {journal} {\bibinfo  {journal} {Phys. Rev. B}\ }\textbf {\bibinfo {volume} {103}},\ \bibinfo {pages} {085117} (\bibinfo {year} {2021}{\natexlab{a}})}\BibitemShut {NoStop}%
\bibitem [{\citenamefont {Ogino}\ \emph {et~al.}(2021{\natexlab{b}})\citenamefont {Ogino}, \citenamefont {Furukawa}, \citenamefont {Kaneko}, \citenamefont {Morita},\ and\ \citenamefont {Kawashima}}]{oginoSymmetryprotectedTopologicalPhases2021}%
  \BibitemOpen
  \bibfield  {author} {\bibinfo {author} {\bibfnamefont {T.}~\bibnamefont {Ogino}}, \bibinfo {author} {\bibfnamefont {S.}~\bibnamefont {Furukawa}}, \bibinfo {author} {\bibfnamefont {R.}~\bibnamefont {Kaneko}}, \bibinfo {author} {\bibfnamefont {S.}~\bibnamefont {Morita}},\ and\ \bibinfo {author} {\bibfnamefont {N.}~\bibnamefont {Kawashima}},\ }\bibfield  {title} {\bibinfo {title} {Symmetry-protected topological phases and competing orders in a spin-1/2 {{XXZ}} ladder with a four-spin interaction},\ }\href {https://doi.org/10.1103/PhysRevB.104.075135} {\bibfield  {journal} {\bibinfo  {journal} {Phys. Rev. B}\ }\textbf {\bibinfo {volume} {104}},\ \bibinfo {pages} {075135} (\bibinfo {year} {2021}{\natexlab{b}})}\BibitemShut {NoStop}%
\bibitem [{\citenamefont {Fontaine}\ \emph {et~al.}(2024)\citenamefont {Fontaine}, \citenamefont {Sugimoto},\ and\ \citenamefont {Furukawa}}]{fontaineSymmetryTopologyDuality2024}%
  \BibitemOpen
  \bibfield  {author} {\bibinfo {author} {\bibfnamefont {M.}~\bibnamefont {Fontaine}}, \bibinfo {author} {\bibfnamefont {K.}~\bibnamefont {Sugimoto}},\ and\ \bibinfo {author} {\bibfnamefont {S.}~\bibnamefont {Furukawa}},\ }\bibfield  {title} {\bibinfo {title} {Symmetry, topology, duality, chirality, and criticality in a spin-1/2 {{XXZ}} ladder with a four-spin interaction},\ }\href {https://doi.org/10.1103/PhysRevB.109.134413} {\bibfield  {journal} {\bibinfo  {journal} {Phys. Rev. B}\ }\textbf {\bibinfo {volume} {109}},\ \bibinfo {pages} {134413} (\bibinfo {year} {2024})}\BibitemShut {NoStop}%
\bibitem [{\citenamefont {Weyrauch}\ and\ \citenamefont {Rakov}(2017)}]{weyrauch}%
  \BibitemOpen
  \bibfield  {author} {\bibinfo {author} {\bibfnamefont {M.}~\bibnamefont {Weyrauch}}\ and\ \bibinfo {author} {\bibfnamefont {M.~V.}\ \bibnamefont {Rakov}},\ }\bibfield  {title} {\bibinfo {title} {Dimerization in ultracold spinor gases with zeeman splitting},\ }\href {https://doi.org/10.1103/PhysRevB.96.134404} {\bibfield  {journal} {\bibinfo  {journal} {Phys. Rev. B}\ }\textbf {\bibinfo {volume} {96}},\ \bibinfo {pages} {134404} (\bibinfo {year} {2017})}\BibitemShut {NoStop}%
\bibitem [{\citenamefont {Roberts}\ \emph {et~al.}(2019)\citenamefont {Roberts}, \citenamefont {Jiang},\ and\ \citenamefont {Motrunich}}]{Roberts_2019}%
  \BibitemOpen
  \bibfield  {author} {\bibinfo {author} {\bibfnamefont {B.}~\bibnamefont {Roberts}}, \bibinfo {author} {\bibfnamefont {S.}~\bibnamefont {Jiang}},\ and\ \bibinfo {author} {\bibfnamefont {O.~I.}\ \bibnamefont {Motrunich}},\ }\bibfield  {title} {\bibinfo {title} {Deconfined quantum critical point in one dimension},\ }\bibfield  {journal} {\bibinfo  {journal} {Physical Review B}\ }\textbf {\bibinfo {volume} {99}},\ \href {https://doi.org/10.1103/physrevb.99.165143} {10.1103/physrevb.99.165143} (\bibinfo {year} {2019})\BibitemShut {NoStop}%
\bibitem [{\citenamefont {Regnault}\ \emph {et~al.}(1994)\citenamefont {Regnault}, \citenamefont {Zaliznyak}, \citenamefont {Renard},\ and\ \citenamefont {Vettier}}]{PhysRevB.50.9174}%
  \BibitemOpen
  \bibfield  {author} {\bibinfo {author} {\bibfnamefont {L.~P.}\ \bibnamefont {Regnault}}, \bibinfo {author} {\bibfnamefont {I.}~\bibnamefont {Zaliznyak}}, \bibinfo {author} {\bibfnamefont {J.~P.}\ \bibnamefont {Renard}},\ and\ \bibinfo {author} {\bibfnamefont {C.}~\bibnamefont {Vettier}},\ }\bibfield  {title} {\bibinfo {title} {Inelastic-neutron-scattering study of the spin dynamics in the haldane-gap system {Ni(C$_2$H$_8$N$_2$)NO$_2$ClO$_4$}},\ }\href {https://doi.org/10.1103/PhysRevB.50.9174} {\bibfield  {journal} {\bibinfo  {journal} {Phys. Rev. B}\ }\textbf {\bibinfo {volume} {50}},\ \bibinfo {pages} {9174} (\bibinfo {year} {1994})}\BibitemShut {NoStop}%
\bibitem [{\citenamefont {Zheludev}\ \emph {et~al.}(1996)\citenamefont {Zheludev}, \citenamefont {Nagler}, \citenamefont {Shapiro}, \citenamefont {Chou}, \citenamefont {Talham},\ and\ \citenamefont {Meisel}}]{PhysRevB.53.15004}%
  \BibitemOpen
  \bibfield  {author} {\bibinfo {author} {\bibfnamefont {A.}~\bibnamefont {Zheludev}}, \bibinfo {author} {\bibfnamefont {S.~E.}\ \bibnamefont {Nagler}}, \bibinfo {author} {\bibfnamefont {S.~M.}\ \bibnamefont {Shapiro}}, \bibinfo {author} {\bibfnamefont {L.~K.}\ \bibnamefont {Chou}}, \bibinfo {author} {\bibfnamefont {D.~R.}\ \bibnamefont {Talham}},\ and\ \bibinfo {author} {\bibfnamefont {M.~W.}\ \bibnamefont {Meisel}},\ }\bibfield  {title} {\bibinfo {title} {Spin dynamics in the linear-chain {$S=1$} antiferromagnet {Ni}({C}$_3${H}$_{10}${N}$_2$)$_2${N}$_3$({ClO}$_4$)},\ }\href {https://doi.org/10.1103/PhysRevB.53.15004} {\bibfield  {journal} {\bibinfo  {journal} {Phys. Rev. B}\ }\textbf {\bibinfo {volume} {53}},\ \bibinfo {pages} {15004} (\bibinfo {year} {1996})}\BibitemShut {NoStop}%
\bibitem [{\citenamefont {Zheludev}\ \emph {et~al.}(2001)\citenamefont {Zheludev}, \citenamefont {Chen}, \citenamefont {Broholm}, \citenamefont {Honda},\ and\ \citenamefont {Katsumata}}]{PhysRevB.63.104410}%
  \BibitemOpen
  \bibfield  {author} {\bibinfo {author} {\bibfnamefont {A.}~\bibnamefont {Zheludev}}, \bibinfo {author} {\bibfnamefont {Y.}~\bibnamefont {Chen}}, \bibinfo {author} {\bibfnamefont {C.~L.}\ \bibnamefont {Broholm}}, \bibinfo {author} {\bibfnamefont {Z.}~\bibnamefont {Honda}},\ and\ \bibinfo {author} {\bibfnamefont {K.}~\bibnamefont {Katsumata}},\ }\bibfield  {title} {\bibinfo {title} {Haldane-gap excitations in the low-${H}_{c}$ one-dimensional quantum antiferromagnet {Ni}({C}$_{5}${D}$_{14}${N}$_{2}$)$_{2}${N}$_{3}(${PF}$_{6}$)},\ }\href {https://doi.org/10.1103/PhysRevB.63.104410} {\bibfield  {journal} {\bibinfo  {journal} {Phys. Rev. B}\ }\textbf {\bibinfo {volume} {63}},\ \bibinfo {pages} {104410} (\bibinfo {year} {2001})}\BibitemShut {NoStop}%
\bibitem [{\citenamefont {Chepiga}\ and\ \citenamefont {Mila}(2017)}]{PhysRevB.96.054425}%
  \BibitemOpen
  \bibfield  {author} {\bibinfo {author} {\bibfnamefont {N.}~\bibnamefont {Chepiga}}\ and\ \bibinfo {author} {\bibfnamefont {F.}~\bibnamefont {Mila}},\ }\bibfield  {title} {\bibinfo {title} {Excitation spectrum and density matrix renormalization group iterations},\ }\href {https://doi.org/10.1103/PhysRevB.96.054425} {\bibfield  {journal} {\bibinfo  {journal} {Phys. Rev. B}\ }\textbf {\bibinfo {volume} {96}},\ \bibinfo {pages} {054425} (\bibinfo {year} {2017})}\BibitemShut {NoStop}%
\end{thebibliography}%

\end{document}